\documentclass{aa}
\usepackage{graphicx}

\newcommand{\Teff}{\mbox{$T_{\rm eff} $}}
\newcommand{\logg}{\mbox{$\log g $}}

\newcommand{\DanCor}{{GSC2U\,J131147.2+292348 }}

\begin{document}

\title{Model atmosphere analysis of the extreme DQ white dwarf \DanCor}


\author{D. Carollo,
    \inst{1}
    D. Koester,
    \inst{2}
    A. Spagna,
      \inst{1}
    M.G. Lattanzi,
      \inst{1}
    S.T. Hodgkin
      \inst{3}
        }

 \institute
   {INAF, Osservatorio Astronomico di
   Torino, I-10025 Pino Torinese, Italy
    \and Institut f\"ur Theoretische Physik und Astrophysik,
   Universit\"at Kiel, 24098 Kiel, Germany
   \and Cambridge Astronomical Survey Unit, Institute of
    Astronomy, Madingley Road, Cambridge, CB3 0HA, UK\\
    }

\date{Received 09 January 2003/Accepted 29 January 2003}

\abstract{A new model atmosphere analysis for the peculiar DQ
white dwarf discovered by Carollo et al.\ (2002) is presented. The
effective temperature and carbon abundance have been estimated by
fitting both the photometric data (UB$_{\rm J}$VR$_{\rm F}$I$_{\rm
N}$JHK) and a low resolution spectrum ($3500<\lambda<7500$ \AA)
with a new model grid for helium-rich white dwarfs with traces of
carbon (DQ stars). We estimate $T_{\rm eff}\simeq 5120\pm 200$ K
and log[C/He]$\simeq -5.8 \pm 0.5$, which make
GSC2U\,J131147.2+292348 the coolest DQ star ever observed.  This
result indicates that the hypothetical transition from C$_2$ to
C$_2$H molecules around \Teff\ = 6000~K, which was inferred to
explain the absence of DQ stars at lower temperatures, needs to be
reconsidered. \keywords{White dwarfs -- Techniques: spectroscopic
-- Stars: kinematics -- Stars: individual(GSC2U J131147.2+292348)}
}

\titlerunning{\DanCor}

\maketitle


\section{Introduction}

The presence of carbon in the atmospheres of some non-DA white
dwarfs (defined as spectral type DQ) is generally explained by the
convective dredge-up from the stellar core to the outer
photospheric layers (Koester et al.\ 1982; Pelletier et al.\
1986).  C$_2$ molecules are responsible for the absorption bands
(e.g.\ in particular the Swan bands) which are the typical
signature of the DQ stars. The spectral energy distribution of
these stars changes significantly as a function of the effective
temperature, $T_{\rm eff}$, and carbon abundance, [C/He], as shown
by the theoretical atmosphere models of Koester et al.\ (1982) and
Wegner \& Yackovich (1984).  Typically, strong absorption bands
are expected for the coolest DQ stars, even in the case of low
carbon abundances.


However, past surveys revealed DQ stars with effective temperature
above 6500 K only (Bergeron et al.\ 1997). The existence of this
cut-off is not well understood. In fact, if cool DQ stars with
$T_{\rm eff}<6500$ K do exist, their strong Swan bands  would
result in peculiar colors and spectra which should make these
objects easily recognizable.

On the other hand, at low temperature carbon can be present also
in a different form, as C$_2$H molecules, if some hydrogen is also
present. The electronic transition spectra of the C$_2$H are not
known from theory or laboratory experiments, but the observed
spectra of the few known C$_2$H stars show molecular absorption
bands similar to the Swan bands shifted by about 150 {\AA} toward
the blue. This shift cannot be explained as an effect of pressure
shift of the Swan bands in a helium dominated atmosphere (Bergeron
at al.\ 1994) or as a displacement due to a magnetic field
(Schmidt et al.\ 1995).  The presence of a certain fraction of
hydrogen in the atmosphere of such non-DA white dwarfs can also be
inferred by the collision induced absorption in the near IR due to
H$_2$ molecules, as in the case of LHS 1126.

These observations suggest the hypothesis that DQ white dwarfs
turn into C$_2$H stars when $T_{\rm eff}$ is below 6500 K, due to
a not well identified physical mechanism that should inject
hydrogen\footnote {The evolution of the ``missing'' cool DQ stars
is related to the more general problem of the non-DA gap which
derives from the apparent lack of non-DA stars observed with
temperature $5100 \la T_{\rm eff}\la 6100$ K. At the moment, the
cause of this effect, which could depend on the physical and
chemical evolution of the white dwarf atmospheres during the
cooling phases as well as on not sufficiently understood input
physics, is not well established (see e.g.\ Bergeron et al.\ 1997,
Malo et al.\ 1999).}  in the He-dominated atmosphere of the DQ
stars (Bergeron et al.\ 1997, 2001).

Recently, Carollo et al.\ (2002) discovered \DanCor during a
proper motion survey for halo white dwarfs based on the
photographic material used for the construction of the GSC-II
(McLean et al.\ 2000). As shown in Figure~2 of Carollo et al.\
(2002), this object appears as a very peculiar carbon rich white
dwarf due to the simultaneous presence of strong C$_2$
Deslandres-d'Azambuja and Swan bands, with an evident depression
of the continuum in the Swan region between 4500 and 6200 {\AA}.
No other DQ star shows both these extreme features.


\DanCor represents an enigma as well as an opportunity to test the
predictions of the scenario discussed above for these objects.  In
fact, temperatures as low as 6000~K were already estimated by
Carollo et al.\ (2002) by means of a simple spectral analysis. As
described in the next sections, here we adopt a more sophisticated
fitting technique based on new atmosphere models in order to
estimate accurately the temperature and composition of this
object. The new results confirm cool values of the temperature, $
T_{\rm eff}\simeq 5100$ K, and the implications of this result
will be briefly discussed.

\begin{table*}
\centering \caption{Observed standard and photographic magnitudes
for \DanCor\ and theoretical fits.}
\begin{tabular}{lrrrrrrrrr}
\hline\hline observations & \Teff/$\log$[C/He] & U  & B$_J$  & V &
R$_F$ & I$_N$ & J & H & K \\
UBVRIJHK & &19.15&19.60&19.10&18.10&17.50&17.48&17.13&17.08\\
errors        & & 0.15& 0.15& 0.15& 0.15& 0.15& 0.05& 0.10& 0.12\\
model    &4980/-6.17&19.06&19.59&19.11&18.19&17.81&17.44&17.18&17.02\\
\hline
UVJHK    & &19.15& &19.10& &&17.48&17.13&17.08\\
model    & 4955/-6.32&19.15& &19.10& & &17.47&17.20&17.04\\
\hline
\end{tabular}

\label{magres}
\end{table*}

\begin{figure}

\includegraphics[width=8cm]{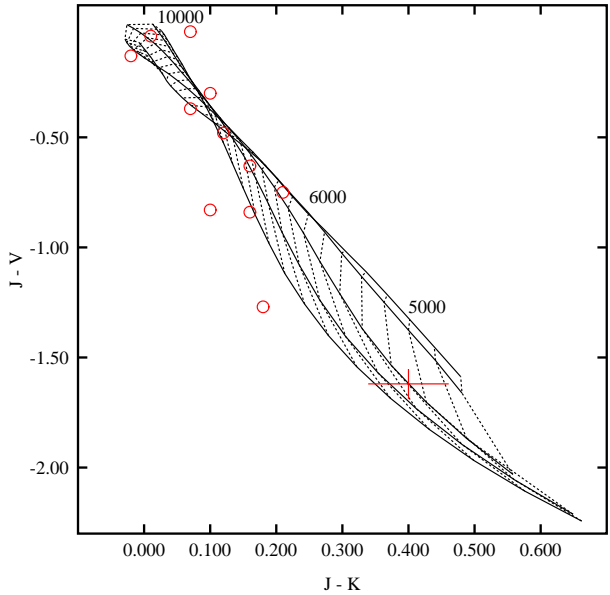}
\caption{Two-color diagram constructed from $VJK$. For the
theoretical predictions the continuous lines
  are lines of constant carbon abundance from $\log$[C/He] = $-8$ (top) to $-4$ (bottom);
  dotted lines are lines of
  constant effective temperature as described in the text.
The cross marks \DanCor which is the coolest object near 5100~K,
while the circles indicate other DQ stars.}
\label{twocol}
\end{figure}

\begin{figure*}

\includegraphics[width=\textwidth]{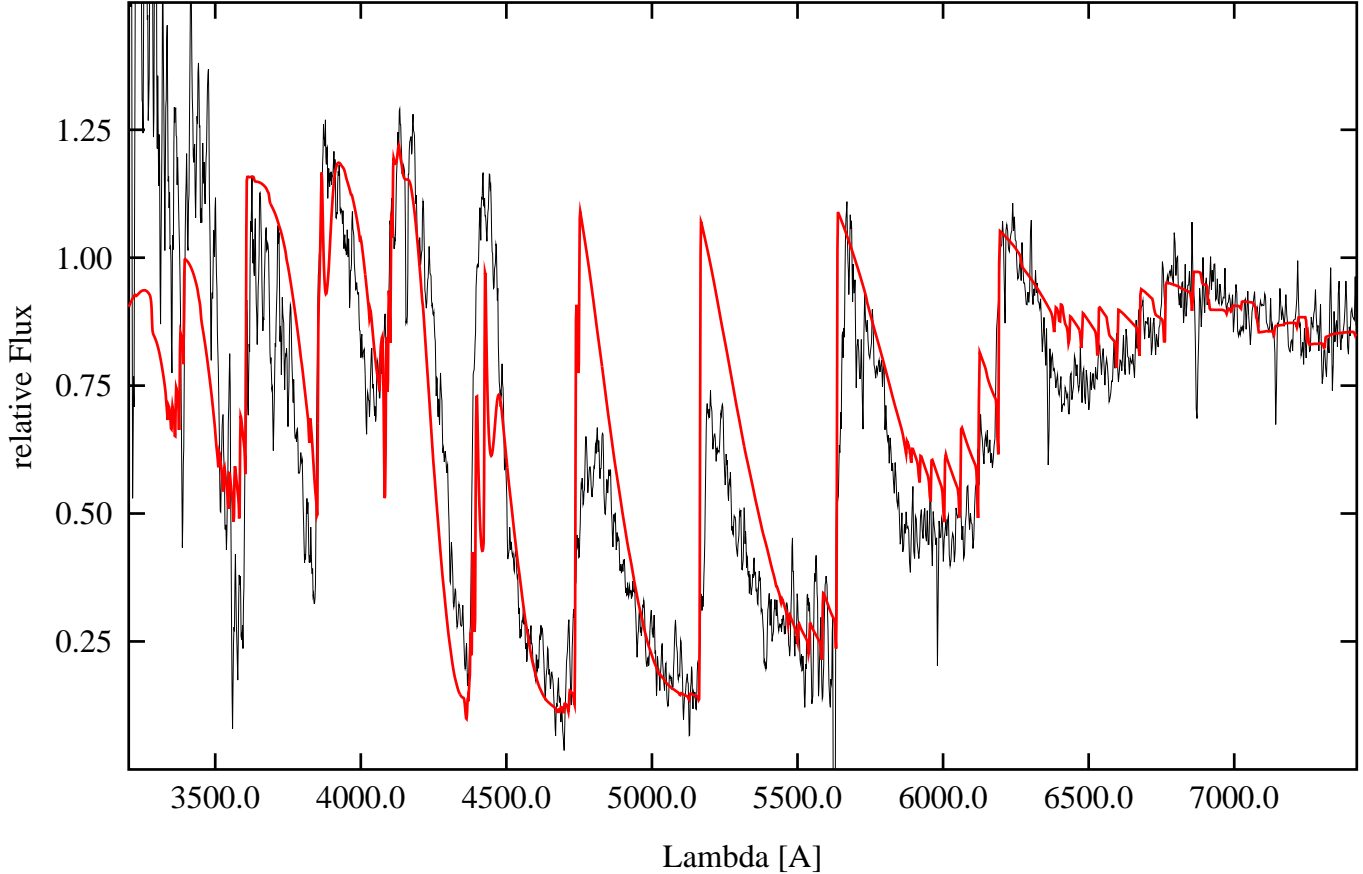}
\caption{Observed spectrum of \DanCor (thin line) and theoretical
  model (thick line).}
 \label{specfit}
\end{figure*}

\section{Model atmosphere analysis of \DanCor}
\DanCor\ shows extremely strong bands of the C$_2$ molecule,
especially the Swan and Deslandres-d'Azambuja band systems in the
optical range. As these bands obviously block a significant
fraction of the total flux, they will influence the temperature
structure of the atmosphere models. We have therefore calculated a
new grid of model atmospheres, which takes into account the
blanketing effect of 537 bands from the Swan,
Deslandres-d'Azambuja, Fox-Herzberg, Phillips, and Ballik-Ramsay
systems, as well as numerous atomic carbon lines and the resonance
lines of HeI in the EUV. The general molecular data were taken
from Huber \& Herzberg (1979), the Franck-Condon factors for the
vibrational transitions from various sources (Krishnaswamy \&
Odell 1977; Dwivedi et al. 1978; Spindler 1965; Sharp 1983).
Molecular absorption is treated with the ``just overlapping line
approximation (JOLA)'' in the version as described in Zeidler-K.T.
\& Koester (1982).

The general procedures and input physics of the model atmosphere
calculations are very similar to the description in Finley et al.
(1997). As it is practically impossible for very cool white dwarfs
to determine effective temperature, surface gravity, and in our
case the carbon abundance simultaneously, we have held \logg\
fixed at the canonical value of 8.00. \Teff\ for the grid ranges
from 10000 to 4600~K, the abundance ratio $\log$[C/He] by numbers
from $-8$ to $-4$.

\subsection{Magnitudes and colors}
For cool white dwarfs magnitudes and colors, especially in the
infrared, are very useful for the determination of atmospheric
parameters, a method pioneered by Bergeron et al. (1997).

Since magnitudes in both the standard and photographic system,
with a spectral coverage from the ultraviolet to the near IR, have
been observed for \DanCor, we have also calculated theoretical
magnitudes for our model grid. We adopted U,V from the
photographic photometry given by Moreau \& Reboul (1995), while
Carollo et al. (2002) provided photographic B$_J$, R$_F$ and I$_N$
in the natural photographic system of the POSS-II plates, plus
standard JHK photometry from observations carried out at the 4-m
TNG (La Palma). The methods and the magnitude zeropoints used for
the UV-JHK bandpasses in the standard Johnson system are described
in detail in Zuckerman et al. (2003). For the photographic B$_J$,
R$_F$ and I$_N$ (approximately corresponding to the
Johnson-Cousins B(RI)$_c$) we have used  the same transmission
curves adopted for the photometric calibration of the GSC-II
plates and determined the zeropoints from integrations over the
Vega flux as obtained from the STScI archive.


The available magnitudes from $U$ to $K$ completely determine the
energy distribution of \DanCor. We have used our automatic least
squares fitting routine, described in Zuckerman et al. (2003) to
determine the best fitting parameters within the \Teff\ -
$\log$[C/He] grid, resulting in an extremely low effective
temperature around 5000~K (Table~\ref{magres}).  The first row in
the Table gives the observed magnitudes, the second the assumed
errors. The third row are the theoretical predictions for the best
fit parameters \Teff\ = 4980~K, $\log$[C/He] = $-6.17$.

As an internal check, we tested the effect of fitting the physical
parameters with only the standard UVJHK photometry with respect to
the global solution including also the B$_J$, R$_F$ and I$_N$
magnitudes. The last two rows of Table~\ref{magres} show that the
parameters change only very little, indicating that the GSC-II
magnitudes are certainly very consistent with the overall energy
distribution.


Figure~\ref{twocol} shows the position of \DanCor\ in a special
two-color diagram $J-V$ vs. $J-K$, using only standard magnitudes
to be able to compare with other known DQ white dwarfs. Continuous
lines are lines of constant carbon abundance, from -4.0 to -8.0,
dotted lines are lines of constant effective temperature from
4600~K to 10000~K in steps of 200~K. As can be seen, this diagram
is not very useful at temperatures above 7000~K, because of the
competing direct effect of flux blocking and the indirect
blanketing effect on the temperature structure. However, in the
range 4600 - 6600~K and the abundances considered here, the
diagram gives a clear indication of the atmospheric parameters.
The cross at the lowest temperatures is \DanCor, for which we
would determine \Teff = 5100, $\log$[C/He] = $-6.0$ from this
position. The other 11 circles are observations of DQ white dwarfs
from Bergeron et al. (1997) and Bergeron et al. (2001), which
clearly are all much hotter, in agreement with temperatures
derived in Bergeron et al. (1997).

\subsection{Spectral fitting}

The spectrum of \DanCor\ has been described in detail in Carollo
et al.\ (2002). They concluded that the extremely strong bands of
the Swan and Deslandres-d'Azambuja systems in the optical range
are compatible with models calculated in Wegner \& Yackovich
(1984), whereas the energy distribution in the infrared could be
explained by a blackbody distribution of around 6000~K. With our
consistent model atmospheres available, we can apply our standard
spectral fitting technique (e.g. Koester et al. 2001) with a
Levenberg-Marquard algorithm (Press et al. 1992) to find the
minimum $\chi^2$ solution, using \Teff\ and $\log$[C/He] as two
free fitting parameters instead of the usual \Teff\ and \logg\ in
the case of DA or DB white dwarfs. The quasi continuum was forced
to fit the model at two positions (around 4150 and 7000~\AA),
allowing for remaining small calibration errors of the spectral
flux. The resulting parameters for the best fit are \Teff =
5200~K, $\log$[C/He] = $-5.53$. Figure 2 shows the observed
spectrum together with the theoretical model corresponding to
these parameters. Qualitatively, the theoretical model describes
the main features of the spectrum, in particular the very strong
band systems. In the details discrepancies remain, which may have
a number of origins: the temperature structure of the models, the
equation of state in these very high pressure atmospheres, and,
most likely, missing bands, due to unkown Franck-Condon factors
for the bands with highly excited lower levels, which are weak at
laboratory conditions, but may be important in the much hotter
stellar atmosphere. Nevertheless, we consider the fit satisfactory
and a confirmation of the low temperatures derived from the
photometry.

\subsection{Results and conclusion}
The fitting procedure for the photometry as well as for the
spectrum provides formal errors, derived from the assumed
statistical errors of the observations. These are very small ---
typically 30-40~K for \Teff\ and 0.05 for $\log$[C/He] --- and
definitely unrealistic, because the errors are dominated by
systematic errors of the models and reductions. These errors can
be estimated only very roughly, taking the differences between the
solutions from photometry and spectrum as a guide. Since we
believe that the spectral result is more reliable, we give it
double weight and take as the final result for the atmospheric
parameters \Teff\ = 5120$\pm$200~K and $\log$[C/He] =
-5.8$\pm$0.5. The distance modulus obtained from the photometric
solution is 3.69 mag, corresponding to a distance of 55~pc and to
a tangential velocity $V_{\rm tan}=4.74 \cdot \mu \, d \simeq 125$
~km~s$^{-1}$ ($\mu=0.48''$ yr$^{-1}$). Adopting the same
kinematics assumptions as in Carollo et al.\ (2002), we obtain
galactic velocity components with respect to the LRS, $(U,V)\simeq
(-115, -1)$ km~s$^{-1}$. These values are well consistent with the
velocity ellipsoid of the galactic halo (1$\sigma$) and are still
consistent with the thick disk kinematics (2$\sigma$), while the
membership of \DanCor to the thin disk appears much less probable.

However, one needs to keep in mind that the results have been
derived using a fixed surface gravity of \logg = 8.00. While we do
not expect the atmospheric parameters to change much with \logg,
the distance modulus depends of course on the radius of the star,
which depends strongly on the assumed surface gravity. Allowing
for a plausible range of 7.5 - 8.5, the radius could be different
up to $\pm$30\%, with the same change resulting for the distance
and velocity.


With \Teff\ about 5100~K this star is by far the coolest known
``normal'' DQ object. It is below the cutoff seen by Bergeron et al.\
(1997) near 6500~K and also below or at least at the lower edge of the
so-called non-DA gap. It cannot be true therefore that all DQ turn
into C$_2$H stars when they cool down, and one obvious explanation
could be that some stars completely avoid any accretion of hydrogen,
which is a prerequisite for the formation of this molecule. However,
the final explanation of this puzzle as well as others concerning the
non-DA gap will likely need to wait for the discovery of more similar
objects from the ongoing large scale survey like the SDSS.

\begin{acknowledgements}
 The authors are grateful to B.J. M$^{c}$Lean for his constant support of this program.
 The GSC II is a joint project of the Space Telescope
Science Institute and the Osservatorio Astronomico di Torino.
Space Telescope Science Institute is operated by AURA for NASA
under contract NAS5-26555. Partial financial support to this
research comes from the Italian CNAA and the Italian Ministry of
Research (MIUR) through the COFIN-2001 program.

This work is based on observations made with the Italian
Telescopio Nazionale Galileo (TNG) operated on the island of La
Palma by the Centro Galileo Galilei of the INAF (Istituto
Nazionale di Astrofisica) at the Spanish Observatorio del Roque de
los Muchachos of the Instituto de Astrofisica de Canarias, and
also based on observations made with the William Herschel
Telescope operated on the island of La Palma by the Isaac Newton
Group in the Spanish Observatorio del Roque de los Muchachos of
the Instituto de Astrofisica de Canarias.

\end{acknowledgements}


\begin{thebibliography}{}

\bibitem[\protect\astroncite{{Bergeron} et~al.}{2001}]{Bergeron.Leggett.ea01}
{Bergeron} P., {Leggett} S.~K., {Ruiz} M., 2001,
  ApJS, 133, 413

\bibitem[\protect\astroncite{{Bergeron} et~al.}{1997}]{Bergeron.Ruiz.ea97}
{Bergeron} P., {Ruiz} M.~T., {Leggett} S.~K., 1997,
  ApJS,  108, 339

\bibitem[\protect\astroncite{{Carollo} et~al.}{2002}]{Carollo.Hodgkin.ea02}
{Carollo}, D., {Hodgkin}, S. T., {Spagna}, A., {Smart}, R. L.,
{Lattanzi}, M. G., {McLean}, B. J., {Pinfield}, D. J. 2002, A\&A
393, 45

\bibitem[\protect\astroncite{Dwivedi
et~al.}{1978}]{Dwivedi.Branch.ea78} Dwivedi, P.H., Branch, D.,
Huffaker, J.N. 1978, ApJS, 36, 573

\bibitem[\protect\astroncite{{Finley} et~al.}{1997}]{Finley.Koester.ea97}
{Finley} D.~S., {Koester} D., {Basri} G., 1997,
  ApJ  488, 375

\bibitem[\protect\astroncite{Huber \&
  Herzberg}{1979}]{Huber.Herzberg79} Huber, K.P., Herzberg, G. 1979,
  {Constants of Diatomic Molecules} (New York: Van Nostrand)

\bibitem[\protect\astroncite{{Koester} et~al.}{2001}]{Koester.Napiwotzki.ea01}
{Koester} D., {Napiwotzki} R., {Christlieb} N., {Drechsel} H., {Hagen} H.-J.,
  {Heber} U., {Homeier} D., {Karl} C., {Leibundgut} B., {Moehler} S.,
  {Nelemans} G., {Pauli} E.-M., {Reimers} D., {Renzini} A., {Yungelson} L.,
  2001, AAP~  378, 556

\bibitem[\protect\astroncite{{Koester} et~al.}{1982}]{Koester.ea.82}
{Koester} D.~S., Weidemann, V., Zeidler-K.T. E.-M., 1982,
  A\&A, 116, 147


\bibitem[\protect\astroncite{Krishna Swamy \&
O'Dell}{1977}]{Krishnaswamy.Odell77} {Krishna Swamy}, K.S.,
{O'Dell}, C.R. 1977, ApJ, 216, 158

\bibitem[\protect\astroncite{moreau}{1977}]{moreau.95} {Moreau}, O., Reboul H.  1995,
      A\&A SS, 111, 169

\bibitem[\protect\astroncite{{Press} et~al.}{1992}]{Press.Teukolsky.ea92}
{Press} W.~H., {Teukolsky} S.~A., {Vetterling} W.~T., {Flannery} B.~P., 1992,
\newblock Numerical recipes in FORTRAN. The art of scientific computing,
\newblock Cambridge: University Press, 2nd ed.

\bibitem[\protect\astroncite{Schmidt}{1995}]{Schmidt} Schmidt, G.
D., Bergeron, P., Fegley B.. Jr, 1995, ApJ, 443, 274


\bibitem[\protect\astroncite{Sharp}{1983}]{Sharp83} Sharp, C.M. 1983,
AAS, 55, 33

\bibitem[\protect\astroncite{Spindler}{1965}]{Spindler65} Spindler,
R.J. 1965, JQSRT, 5, 165

\bibitem[\protect\astroncite{{Wegner} \&
  {Yackovich}}{1984}]{Wegner.Yackovich84}
{Wegner} G., {Yackovich} F.~H., 1984,
  ApJ,  284, 257

\bibitem[\protect\astroncite{Zeidler-K.T. \&
Koester}{1982}]{Zeidler.Koester82} Zeidler-K.T., E.-M., Koester,
D. 1982, A\&A, 113, 173



\bibitem[\protect\astroncite{Zuckerman
  et~al.}{2003}]{Zuckerman.Koester.ea03} Zuckerman, B., Koester, D.,
  Reid, I.N., H\"unsch, M. 2003, in preparation

\end{thebibliography}
\end{document}